\documentclass[
  ,draft            
  ,sort&compress
  ]
  {aipproc}

\layoutstyle{8x11double}

\def\la{\;\raise0.3ex\hbox{$<$\kern-0.75em\raise-1.1ex\hbox{$\sim$}}\;}
\def\ga{\;\raise0.3ex\hbox{$>$\kern-0.75em\raise-1.1ex\hbox{$\sim$}}\;}
\newcommand{\msun}{\mbox{$M_\odot$}}

\begin{document}

\title{Theory of cooling neutron stars versus observations}

\classification{26.60.-c; 26.60.Dd; 97.60.Jd}
\keywords      {neutron stars; thermal emission; neutrino emission}

\author{D.~G.\ Yakovlev}{
  address={Ioffe Physico-Technical Institute, Politekhnicheskaya 26,
  194021, St.\ Petersburg, Russia}
   ,altaddress={Joint Institute for Nuclear Astrophysics,
   210a Nieuwland Science Hall,
   Notre Dame, IN 46556-5670, USA} 
}

\author{O.~Y.\ Gnedin}{
  address={University of Michigan,
    500 Church Street, Ann Arbor, MI 48109-1042, USA}
}

\author{A.~D.\ Kaminker}{
  address={Ioffe Physico-Technical Institute, Politekhnicheskaya 26,
  194021, St.\ Petersburg, Russia}
}

\author{A.~Y.\ Potekhin}{
  address={Ioffe Physico-Technical Institute, Politekhnicheskaya 26,
  194021, St.\ Petersburg, Russia}
   ,altaddress={Ecole Normale Sup\'erieure de Lyon, CRAL (UMR CNRS 5574),
   46 all\'ee d'Italie,
   Lyon 69364, France} 
}

\begin{abstract}
We review current state of neutron star cooling
theory and discuss the prospects to constrain
the equation of state,
neutrino emission and superfluid properties of 
neutron star cores
by comparing the cooling theory with observations
of thermal radiation from isolated neutron stars.
\end{abstract}


\maketitle

`

\section{Introduction}

The equation of state (EOS) of superdense matter
in neutron star cores is still a mystery.
It is not clear if it is soft, moderate or stiff;
if the matter contains nucleons/hyperons, or
exotic components. In the absence of good practical
theory of supranuclear matter
the problem cannot be solved on purely theoretical basis,
but it can be solved by comparing theoretical models
with observations of neutron stars. 
The attempts to solve
this long-standing problem by different
methods are numerous (e.g., Refs.\ \cite{nsb1,lp07}).
Here we discuss current results obtained
from studies of cooling isolated neutron stars.

The first papers on neutron star cooling 
appeared with an advent of X-ray astronomy,
before the discovery of neutron stars.
Their authors tried to prove that neutron stars 
cool not too fast and can be discovered as
sources of thermal surface X-ray radiation.
The first estimates of thermal emission from cooling
neutron stars were most probably done by Stabler \cite{stabler60}
in 1960. Four years later
Chiu \cite{chiu64}
made similar estimates and analyzed the possibility
to discover neutron stars from their thermal emission.
First, simplified calculations of neutron star cooling
were done in 1964 and 1965
\cite{morton64,
cs64, 
bw65b}.
The foundation of the strict cooling theory was laid
in 1966 by Tsuruta and Cameron
\cite{tc66}, one year before the discovery of pulsars.
We review the current state of the cooling theory.
More details can be found in recent review papers
\cite{yp04,pgw06}.

\section{Cooling neutron stars}

Neutron stars are thought 
to consist of a thin crust
and a massive core (e.g., Ref.\ \cite{nsb1}). The crust constitutes
a few per 
cent
of the star's mass and is about $\sim 1$ km thick.
The mass density $\rho$ at the crust-core boundary
is $\approx \rho_0/2$, where $\rho_0\approx 2.8 \times 10^{14}$
g~cm$^{-3}$ is the density of saturated nuclear matter.   
The crust is further divided into the outer 
and inner crust. The outer crust (more generally, the
outer envelope; where $\rho$ is below the neutron drip density
$\la 4 \times 10^{11}$ g~cm$^{-3}$)
consists of atomic nuclei and strongly degenerate electrons.
In the inner crust, free neutrons appear in addition to
the electrons and nuclei; these neutrons can be in
superfluid state. 

\renewcommand{\arraystretch}{1.2}
\begin{table}[tbh]
\caption{Main slow neutrino emission
in nucleon matter}
  \begin{tabular}{lll}
  \hline  
  Process   &    &  $Q_{\rm s}$ [erg cm$^{-3}$ s$^{-1}$] \\
  \hline 
  Modified Urca &
  ${\rm n+N\to p+N+e+}\bar{\nu}$ \quad
   ${\rm p+N+e\to n+N+}\nu$ &
  $ 10^{20}-3 \times 10^{21}$  \\
  Bremsstrahlung \quad &
  ${\rm N+N\to N+N+}\nu+\bar{\nu}$  &
  $ 10^{19}-10^{20}_{}$\\
   \hline
\end{tabular}
\label{tab-nucore-slow}
\end{table}

The neutron star core can be divided into the outer and
inner core. The outer core extends to $\rho \sim (2-3) \rho_0$
and
consists of neutrons (n), with an admixture of protons (p),
electrons (e) and possibly muons. All these particles are
strongly degenerate. The inner core extends to the stellar
center [to $\sim (10-20)\rho_0$ in most massive stars].
Its composition is very uncertain. It may be the same
as in the outer core or essentially
different. In particular, hyperons may appear there in addition
to nucleons. Another possibility is the appearance
of exotic matter (pion or kaon condensates or strange quark matter
or mixed phases, as reviewed, e.g., in \cite{nsb1}).
Nucleons, hyperons and quarks
can be in superfluid state.
Physical properties
of matter 
at $\rho \la \rho_0$
are
more or less
restricted by nuclear physics data, but
at higher $\rho$ they are uncertain. 
Typical baryon chemical potentials $\sim 500$ MeV
in a neutron star core are most difficult for rigorous
microscopic calculations. 
Note a lively discussion in the literature that compact stars
(or some of them) can be not neutron stars but 
strange stars (e.g., Refs.\ \cite{glendenning96,nsb1,lp07})
built entirely or almost entirely of strange
quark matter.  
  
Neutron stars are born in supernova explosions
with high internal temperature $T \sim 10^{11}$ K,
but gradually cool down. 
In $\sim$30 s after the birth a star becomes 
transparent for neutrinos generated
in its interiors and transforms from 
a proto-neutron star
(e.g., \cite{ponsetal01})
to an ordinary neutron star whose EOS
is almost temperature independent (except near the very
surface). At the
later neutrino-transparent stage
the star cools via
neutrino emission from its interior
and via heat transport to the surface and subsequent
photon thermal emission.

Neutron star cooling is described by general relativistic
equations \cite{thorne77} of heat diffusion inside
the star with neutrino energy sources
and surface photon emission.
The solution gives
the distribution of the temperature $T$ inside the star
versus time $t$, and the effective
surface temperature $T_{\rm s}(t)$. 
To simplify computations, one 
divides \cite{gpe83} the neutron star into the interior region and
the outermost he\-at-blanketing envelope
[extended to densities 
$\rho \le \rho_{\rm b} \sim (10^{10}-10^{11})$ g cm$^{-3}$].
The thermal structure of the blanketing envelope
is studied separately in the stationary,
plane-parallel approximation, that relates
$T_{\rm s}$ to the temperature
$T_{\rm b}$ at $\rho=\rho_{\rm b}$. The diffusion equations
are then solved in the interior ($\rho \geq \rho_{\rm b}$).

The thermal photon luminosity of the star is
$ L_\gamma = 4 \pi \sigma R^2 T^4_{\rm s}(t)$,
where $R$ is the circumferential stellar radius.
Both, $L_\gamma$ and $T_{\rm s}$, refer to a locally-flat
reference frame on the stellar surface. A distant
observer detects the ``apparent'' luminosity
$L_\gamma^\infty = L_\gamma (1 - r_{\rm g}/R)$
and the ``apparent'' effective temperature
$T_{\rm s}^\infty = T_{\rm s} \, \sqrt{1 - r_{\rm g}/R}$,
where $r_{\rm g}=2GM/c^2$ is the Schwarzschild radius
and $M$ is the gravitational neutron star mass.
If the surface temperature distribution is anisotropic
(for instance, owing to a strong magnetic field),
$L_\gamma^\infty$ is determined by a properly
averaged surface temperature (e.g., Ref.\ \cite{py01}).

The theory gives
cooling curves, $T_s^\infty(t)$, and predicts
three main cooling stages.
The initial stage of
internal thermal relaxation lasts $t \la 10-50$ yr.
At this stage, the neutron star crust stays hotter
than the core and thermally decoupled from it
(because of much stronger neutrino cooling in the core). 
The surface temperature is then insensitive
to the physics of the core but strongly
depends on physical properties of the crust
\cite{lattimeretal94,gyp01}.
Since no neutron star has been observed at this
stage, we will not discuss it in detail.
The next stage of neutrino cooling 
with isothermal interior lasts
$t \la 10^5-10^6$ yr. At this stage
the neutrino luminosity
$L_\nu \gg L_\gamma$; thermal conduction
is high and makes the stellar interior isothermal,
with the main temperature gradient located in the
heat blanketing envelope. The cooling is mostly regulated
by a strong neutrino emission from the core. The
surface temperature responds to the core cooling
and depends on properties of superdense core.
All cooling isolated neutron stars whose
thermal surface radiation has been detected seem to be
at this neutrino cooling stage (or its very end).
Therefore, we focus our attention on this stage.
At the next photon cooling stage
the star is cold; its neutrino emission
dies out ($L_\nu \ll L_\gamma$)
and the cooling is governed by photon surface emission.

Soon after  
the cooling  starts 
(in minutes to a year, depending on the internal
structure), the core temperature drops to $T \sim 10^9$~K.
For this $T$, the internal thermal energy 
of the star is $\sim 10^{48}$ erg. 
The cooling theory shows how this heat emerges from the
star.

\section{Neutrino emission mechanisms}

\renewcommand{\arraystretch}{1.2}
\begin{table}
\caption{Leading processes of fast
neutrino emission
in nucleon matter and three models of exotic matter}
  \begin{tabular}{lcc}
  \hline 
 Model              &  Process            &
       $Q_{\rm f}$ [erg cm$^{-3}$ s$^{-1}$] \\
  \hline
 Nucleon matter &
 ${\rm n \to p+ e +\bar{\nu} \quad 
   p +e \to n +\nu }$  & $ 
  10^{26}-3\times 10^{27}$  \\
 Pion condensate &
 ${\rm \widetilde{N} \to \widetilde{N}+ e +\bar{\nu} \quad
   \widetilde{N} +e \to \widetilde{N}+ {\nu} } $  & $ 
  10^{23}-10^{26}$   \\
  Kaon condensate  &
  ${\rm \widetilde{B} \to \widetilde{B}+e +\bar{\nu} \quad
    \widetilde{B} + e \to \widetilde{B} +\nu } $  & $ 
   10^{23}-10^{24}$  \\
   Quark matter &
  ${\rm d \to u +e+ \bar{\nu} \quad  u +e \to d+ \nu } 
   $ & $  
   10^{23}-10^{24}$  \\
   \hline 
\end{tabular}
\label{tab-nucore-fast}
\end{table}

Let us summarize the main neutrino
emission mechanisms in the neutron star core. 
They are strongly affected by baryon superfluidity. 
More details
can be found, e.g., in Refs.\ \cite{pethick92,yls99,ykgh01}.

\paragraph{Neutrino emission in nonsuperfluid cores} 
The major neutrino mechanisms in
nucleon matter of the outer core ($\rho \la 2 \, \rho_0$) are
modified Urca 
process and nucleon-nucleon bremsstrahlung.
They are listed in Table \ref{tab-nucore-slow}
(from Ref.\ \cite{yp04}), where N denotes a nucleon (n or p).
They are relatively weak and produce
slow neutrino cooling. The modified Urca process differs from
its direct Urca progenitor, described below, by an
additional nucleon-spectator that is required to
satisfy momentum conservation of reacting particles.
All reactions involving electrons can involve muons instead
(if available in dense matter).

At higher $\rho$, in the inner core,
neutrino emission can be strongly enhanced by new
mechanisms (Table \ref{tab-nucore-fast}, from \cite{yp04}). 
The enhancement
level greatly depends on the EOS
and composition of superdense matter that is
most important for the cooling problem.
The strongest enhancement is provided by direct Urca process 
\cite{lpph91,pplp92}
in nucleon (or nucleon-hyperon) matter. It is a sequence
of two reactions (a beta decay and beta capture)
forbidden in the outer core by momentum
conservation. It is allowed
in the inner core for the matter with rather
high ($\ga$11--13\%) proton fraction (for EOSs
with large symmetry energy of nuclear matter).
If it is forbidden
but the matter contains pion condensate,  
neutrino emission is enhanced 
(although weaker) by direct-Urca-type
reactions involving quasinucleons $\widetilde{\rm N}$
(superpositions of n and p)
in pion-condensed matter. 
If pion condensate is absent, but kaon condensate
available, neutrino emission is enhanced
(even more weaker) by direct-Urca-like process 
involving baryonic quasiparticles $\widetilde{\rm B}$ in kaon
condensed matter. Nearly the same enhancement
is expected due to the direct Urca process involving d and u
quarks in quark matter.
The processes in the inner core (Table \ref{tab-nucore-fast}) can
amplify the neutrino emission by 2--7 orders of magnitude and
lead to fast cooling. It is also possible 
that the neutrino emission in the inner core is not
enhanced.

The emissivity $Q_\nu(\rho,T)$ of slow and fast neutrino processes
in nonsuperfluid matter
can be written as
\begin{equation}
   Q_{\rm slow}= Q_{\rm s}\, T_9^8,\qquad
   Q_{\rm fast}= Q_{\rm f}\, T_9^6,
\label{Qnu}
\end{equation}
where $T_9=T/(10^9~{\rm K})$, while $Q_{\rm s}$ and $Q_{\rm f}$ are
slowly varying functions of $\rho$ (Tables \ref{tab-nucore-slow} 
and \ref{tab-nucore-fast}).
Thus, the neutrino luminosities of massive 
and low-mass stars (with and without inner core) can be 
very different. For instance,
in a young massive
star, where the direct Urca is open and the core temperature is
$T=10^9$~K, $L_\nu$ is as huge as 
$\sim 10^{46}$ erg~s$^{-1}$.
In a low-mass star at the same $T$, $L_\nu$
would be $\sim 7$ orders of magnitude lower. In both cases
$L_\nu$ rapidly decreases when the star cools.
For instance, one has $L_\nu \sim 10^{34}$ erg~s$^{-1}$
for a low-mass star
at $t\sim10$~kyr. 

\paragraph{Neutrino emission in superfluid cores} 
Nucleons, hyperons, and quarks
in dense matter can be in superfluid state (e.g., \cite{ls01},
also see \cite{yls99,yp04} for references).
This superfluidity occurs via Cooper pairing of particles
owing to an attractive component of their interaction
(with the appearance of a gap
in the particle energy spectrum near the Fermi level).
Systematic simulations of cooling superfluid neutron
stars were triggered by a remarkable paper of Page and
Applegate \cite{pa92}.
Superfluidity of any particles is characterized
by its own density dependent critical temperature $T_{\rm c}(\rho)$.
Microscopic calculations of $T_{\rm c}(\rho)$ are extremely
model dependent and give a large scatter of critical
temperatures. Let us mention several general features.

Neutron pairing in the spin singlet state 
with zero angular momentum ($^1$S$_0$)
occurs in the inner neutron star crust (for free
neutrons) and dies out in the core because singlet-state nuclear
attraction turns into
repulsion near the crust-core interface. Neutron
pairing in the core may occur in a triplet state
with unit angular momentum, $^3$P$_2$ (coupled to
a spin triplet state with three units of angular
momentum, $^3$F$_2$). Cooper pairing of other particles
in the core can occur either in a singlet state or
in a triple state. Baryon superfluidity is affected
by the presence of pion or kaon condensate
(e.g., \cite{tt07} and references therein).
Calculated values $T_{\rm c}(\rho)$
range from $\sim 10^8$~K to a few$\times 10^{10}$~K
and vanish at supranuclear densities (where attractive
pairing interaction becomes
inefficient).

A special case is presented by 
color superconductivity \cite{arw98} owing to 
very strong pairing of unlike quarks in
quark matter, where  
$T_{\rm c}(\rho)$ can be as high as
$\sim 5 \times 10^{11}$~K. 

Any baryon superfluidity
reduces neutrino
processes involving these baryons (Tables \ref{tab-nucore-slow}
and \ref{tab-nucore-fast}) due to
a gap in the baryon energy spectrum. At $T \ll T_{\rm c}$,
the reduction is exponentially strong. For instance,
the powerful direct Urca process can be formally open in
the inner core, but completely suppressed by a strong
superfluidity of neutrons of protons. 

In addition, when $T$ falls below $T_{\rm c}$, superfluidity
initiates a specific neutrino emission 
due to Cooper pairing of baryons \cite{frs76} which
enhances neutrino cooling. This mechanism
can enhance the neutrino luminosity of the star
by a factor of $\la 30-100$ over the 
modified Urca level \cite{pageetal04,gusakovetal04}.

Superfluidity affects also neutron star heat
capacity \cite{yls99} and thermal conductivity
\cite{bhy01,sy07} but 
these effects are less strong.

\section{OBSERVATIONS}

\renewcommand{\arraystretch}{1.2}
\begin{table}[t]   
\caption{Observational limits on surface temperatures of isolated 
neutron stars}
\label{tab-cool-data}
\begin{tabular}{ c l  c  c  c  c  l }
\hline
Number &Source & $t$ [kyr] & $T_{\rm s}^\infty$ [MK] &  Confid.\  & 
Model & Ref.\   \\
\hline
1 &PSR B0531+21 (Crab)& 1      & $<$2.0   & 99.8\%     &  BB &
\cite{weisskopf04}  \\
2 & PSR J0205+6449 (in 3C 58) & 0.82--5.4   & $<$1.02   &  99.8\%  & BB & 
\cite{slane04a}        \\
3 & PSR J1119--6127 & $\sim 1.6$ &  $\approx$ 1.2  &  -- & mHA &
\cite{zavlin07} \\
4 &RX J0822--4300 (in Pup A)   & 2--5   & 1.6--1.9 & 90\% & HA & 
\cite{ztp99}  \\  
%
%
5 & PSR J1357--6429 & $\sim 7.3$ & $\approx 0.766 $ & -- & mHA &
\cite{zavlin07a} \\
6 & RX J0007.0+7303 (in CTA 1) & 10--30 & $<$ 0.66 & -- & BB &
\cite{halpernetal04}\\
7 &PSR B0833--45 (Vela)& 11--25  & $0.68\pm0.03$ & 68\% & mHA &
\cite{pavlovetal01} \\
8 &PSR B1706--44 & $\sim$17  & 0.82$^{+0.01}_{-0.34}$ & 68\% & mHA & 
\cite{mcgowanetal04} \\
9 &PSR J0538+2817 & $30 \pm4$  & $\sim 0.87$ & -- & mHA &
\cite{zp04}  \\
%
10 & PSR B2334+61 & $\sim 41$ & $\sim 0.69 $ & -- & mHA &
\cite{zavlin07} \\
11 &PSR B0656+14      &  $\sim$110 & 0.91$\pm$0.05  & 90\%  & BB &
\cite{possentietal96} \\
12 &PSR~B0633+1748 (Geminga) & $\sim$340 & $\sim 0.5$ & -- & BB &
\cite{kargaltsevetal05} \\
13 &RX~J1856.4--3754    & $\sim$500 & $0.434\pm0.003$ & 68\%  &  ~~mHA$^*$  &
\cite{hoetal07} \\
%
14 &PSR~B1055--52      & $\sim$540 & $\sim 0.75 $& -- & BB &
\cite{pz03}  \\
15 & PSR J2043+2740 & $\sim 1200$ & $\sim 0.44 $ & -- & mHA & 
\cite{zavlin07} \\
16 & RX J0720.4--3125 & $\sim 1300$ & $\sim 0.51$ & -- & ~~HA$^*$ &
\cite{motchetal03}   \\
\hline
\end{tabular}
\end{table}

We will compare cooling theory with observations
of surface thermal radiation of isolated neutron stars.
The current state of the observations is discussed, e.g.,
in Refs.\ \cite{zavlin07,deluca07,kaplan07}.
In Table \ref{tab-cool-data} we list 
16 
isolated neutron
stars whose effective surface temperatures have been
measured or constrained. For brevity, the stars are
numbered, and these numbers are used in the text and figures.

The data include two neutron stars 
(1 and 2, the pulsars Crab and
J0205+6449) in historical supernova remnants (SNRs); 
the famous Vela pulsar 
(7) 
and the similar PSR B1706--44
(8); 
the ``pulsar-twins'' J0538+2817 and B2334+61 (9 and 10);
the ``three musketeers'' [Geminga,  PSR B1055--52,
and PSR B0656+14 (12, 14, and 11)]; one 
compact central object in SNR
(RX J0822--4300, source 4);
the compact source RX J0007.0+7303 at the center
of the SNR CTA 1 
(6);
two ``dim'' (``truly isolated'') neutron stars
RX J1856.4--3754 and RX J0720.4--3125 (13 and 16);
two young and energetic pulsars J1119--6127 
and J1357--6429 (3 and 5); and the old pulsar J2043+2740 (15).

The ages and effective surface temperatures of many sources
are rather uncertain. For the sources mentioned in 
\cite{yakovlevetal04,gusakovetal04,kaminkeretal06a}
the choice of $t$ and $T_{\rm s}^\infty$ is mainly the same
as in these references. The ages of other sources are
pulsar spindown ages, and the errobars of $t$ and $T_{\rm s}^\infty$
are chosen in the same way as in 
\cite{yakovlevetal04,gusakovetal04,kaminkeretal06a}.
We have enlarged a possible age range of the source 2 to 5.4 kyr,
the pulsar spindown age, because this source can be
accidentally projected onto the SNR 3C 58 
(Yu.A.\ Shibanov, private communication, 2007).
The values of $T_{\rm s}^\infty$ are inferred from
observations (reported in the indicated references)
assuming either black-body (BB) spectrum
or 
a 
hydroden atmosphere model 
(magnetic or nonmagnetic one, mHA or HA). 
For the sources 13 and 16,
$T_{\rm s}^\infty$ is determined \cite{hoetal07,motchetal03}
using models of hydrogen atmospheres of finite depth (mHA$^*$
and HA$^*$, respectively).
One important
source, 1E 1207.4--5209, is 
not included in
the table 
because 
of the problems to interpret its spectrum 
(although modeling of such spectra
is progressing \cite{mori_ho}).

\section{Basic cooling curves}

For illustration, we will mainly use models of neutron stars
whose cores contain neutrons, protons and electrons
and have a stiff phenomenological
EOS proposed in Ref.\ \cite{pal88}
(model I for the symmetry energy and the
bulk energy model that gives
the compression modulus of saturated
nuclear matter $K=240$ MeV).
The most massive stable neutron star, for this EOS,
has the gravitational mass $M_{\rm max}=1.977\,\msun$
and the central density $\rho_{\rm c}=2.578 \times 10^{15}$ 
g~cm$^{-3}$;
the direct Urca process is
allowed at 
$\rho \geq \rho_{\rm D}=7.851 \times 10^{14}$ g cm$^{-3}$
(in
stars with  $M \geq M_{\rm D}=1.358\,\msun$).

\begin{figure}
  \includegraphics[width=7.8cm,bb=40 180 570 465]{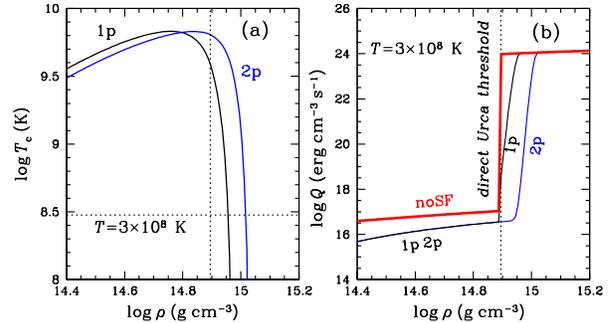}
  \caption{ (a) Superfluid transition temperature 
versus density for two
models (1p and 2p)
of proton superfluidity
in the
neutron star
core. (b) Neutrino
emissivity profiles in the core at $T=3\times10^8$ K
for nonsuperfluid matter (noSF)
and for matter with superfluid protons
(models
1p or 2p).}
\label{tcrho}
\end{figure}

\begin{figure}
  \includegraphics[width=\textwidth, bb=20 15 436 148]{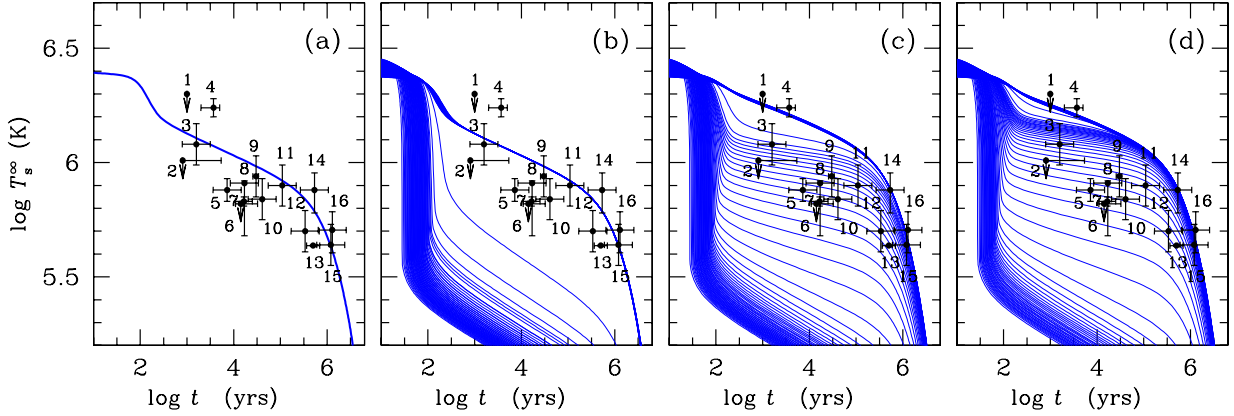}
  \caption{Observational limits on 
  surface temperatures of
  isolated neutron stars compared to
  theoretical cooling curves. (a)~Cooling curve for a nonsuperfluid
  $1.3\,\msun$ star. Any other panel shows 88 curves from
  $M=1.1$ to $1.97\,\msun$ with step $0.01 \msun$:
  (b)~No superfluidity, sharp direct Urca threshold;
  (c)~proton superfluidity 1p and (d)~2p which broaden the
  threshold. }
\label{cool4}
\end{figure}

Neutrino emissivity profiles in a stellar core
at $T=3\times 10^8$~K
are shown in Fig.~\ref{tcrho}b. The vertical dotted
line shows the direct Urca threshold, $\rho_{\rm D}$.
The thick line (NoSF) is for nonsuperfluid matter.
At $\rho<\rho_{\rm D}$ the neutrino emission is
determined by the modified Urca process.
When $\rho$ increases above $\rho_\mathrm{D}$, the
emissivity jumps by 7 orders of magnitude due to
the direct Urca onset.

Fig.\ \ref{cool4}a presents the cooling curve for
a $1.3\msun$ nonsuperfluid star compared with the data. 
The curve is a typical
example of slow cooling via the modified Urca process.
It is actually universal, being 
almost the same for all stars with
$1.1\,\msun \leq M < M_{\rm D}$, and for a wide class
of EOSs \cite{pa92}. This universal curve 
goes through the scatter of observational
points but cannot explain the sources which are hottest
and coolest for their age. These sources
seem to have neutrino luminosities lower and higher
than the modified Urca level.

\begin{figure}
  \includegraphics[width=7.8cm,bb=27 21 312 186]{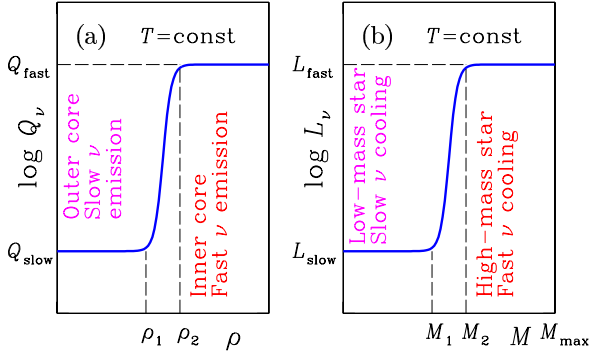}
  \caption{Very schematic view of (a) 
  neutrino emissivity profile in a neutron star core and (b)
  neutrino luminosity versus star's mass at
  a fixed core temperature $T$. Units are arbitrary.}
\label{concept}
\end{figure}

Now consider a set of nonsuperfluid neutron
star models of masses from 1.1 to $1.97\,\msun$, with the mass
step of $0.01\, \msun$ (Fig.\ \ref{cool4}b). 
The first 26 models ($M \leq 1.35\, \msun < M_{\rm D}$)
belong to the class of low-mass stars and show
the universal slow cooling behavior. The next $1.36 \, \msun$ star
has a small central kernel with the enhanced neutrino
emission and shows much faster cooling (is much colder than
all observed sources). All other more massive stars cool
even faster via the direct Urca process and belong
to the family of cold massive rapidly cooling stars.
Evidently, these cooling curves cannot explain the
data.

\section{Basic concept}

A majority of
realistic cooling scenarios, constructed up to now, satisfy 
the following basic phenomenological concept:

(1) Low-mass stars undergo slow neutrino cooling.

(2) Massive stars cool much faster via neutrino emission from their inner
cores.

(3) There is a family of medium-mass stars whose
cooling is intermediate between the slow and fast ones.

Accordingly, the density profile of the neutrino
emissivity in a neutron star core can look like 
that plotted in Fig.\ \ref{concept}a;
$\rho_1$ and $\rho_2$ mark the density range, where
the slow cooling transforms into the fast one.
This $Q_\nu(\rho)$ profile translates into a characteristic 
dependence of
the neutrino luminosity $L_\nu$ on the stellar mass
shown in Fig.\ \ref{concept}b; the transition between
the slow cooling to the fast one takes place in the mass range
from $M_1$ to $M_2$.

Thus, at the present stage of investigation, the cooling
theory \textit{has potential to test four main parameters}
of the neutrino emissivity function $Q_\nu(\rho)$ or 
the luminosity function $L_\nu(M)$. They specify
the lower and upper levels of neutrino emission
(either $Q_\mathrm{slow}$ and $Q_\mathrm{fast}$
or $L_\mathrm{slow}$ and $L_\mathrm{fast}$)
and the position of the transition zone between
the slow and fast emission (either $\rho_1$ and $\rho_2$
or $M_1$ and $M_2$).
 
\section{Examples of physical models}

Two examples \cite{kyg02} of successful realization of the
above scheme are given in Figs.\ \ref{cool4}c and d.
The figures show the cooling curves for neutron stars
of $M=(1.1-1.97)\, \msun$ with strong proton superfluidity
(1p or 2p) in the core (but with normal neutrons).
The density profiles $T_{\rm c}(\rho)$ of the proton critical
temperature are shown in Fig.\ \ref{tcrho}a; they are
phenomenological and can be regarded as illustrative.
In the outer core, proton superfluidity is strong and
suppresses the modified Urca process. Accordingly,
the slow neutrino emission of low-mass stars
is determined by weaker neutrino bremsstrahlung in
neutron-neutron collisions. This rises the surface
temperature of low-mass stars and allows one to explain
observations of hottest sources.

In addition, proton superfluidity penetrates into the
inner core ($\rho > \rho_{\rm D}$) and suppresses
the direct Urca process within the penetration depth.
It broadens the direct Urca threshold and
realizes a smooth transition from slowly to rapidly cooling
neutron stars with the growth of $M$. 
Superfluidity 2p penetrates deeper than 1p which widens
the range of masses of medium-mass stars.
One can now explain
observations of all the sources, including the coldest
ones. Moreover, these scenarios predict the existence
of very cold massive neutron stars (which have not been
observed so far).

If the EOS and the critical temperature
$T_{\rm c}(\rho)$ in the stellar core were known,
one would be able to ``weigh'' neutron stars \cite{kyg02}.
For instance, the mass of the Vela pulsar would be
$1.47\, \msun$ for superfluidity 1p and $1.61\, \msun$ for 2p.
Unfortunately, neither EOS nor $T_{\rm c}(\rho)$
are known and the ``weighing'' procedure is ambiguous.
Nevertheless, as a rule, the ambiguity does not destroy
the mass ordering of cooling stars. For instance,
the Vela pulsar is expected to be more massive than
RX J0822--4300. Moreover, 
one can invert the scheme and state that
had the mass of one or several cooling
neutron stars been measured, 
one could constrain the EOS and $T_{\rm c}(\rho)$. 

One can construct many other models of cooling neutron stars
consistent with the data. One can take other
EOSs in the core [with other direct
Urca thresholds $\rho_{\rm D}$ and functions $T_{\rm c}(\rho)$].
For instance, one can assume the presence of strong neutron
superfluidity and normal protons
(because cooling curves
are nearly symmetric  \cite{gusakovetal04a}
with respect to exchanging neutron and
proton $T_{\rm c}(\rho)$ profiles).

The direct Urca threshold can also be broadened by nuclear
effects (e.g., of pion polarization; see \cite{schaabetal96}
and references therein) or by superstrong magnetic
fields $\ga 3 \times 10^{15}$~G \cite{by99} in a
nonsuperfluid neutron star core. 

\section{Levels of slow and
fast neutrino emission}

\begin{figure}
  \includegraphics[width=\textwidth, bb=10 10 430 140]{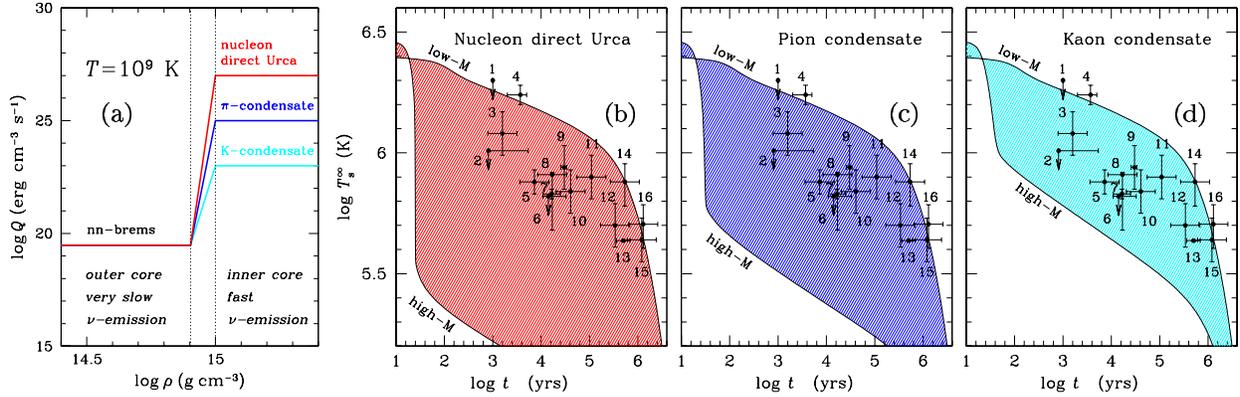}
  \caption{(a) Schematic density dependence of the neutrino emissivity
  in a neutron star core at $T=10^9$~K assuming very slow neutrino
  emission in the outer core and three scenarios of fast
  emission in the inner core. (b), (c), and (d): Ranges of
  $T_{\rm s}^\infty$ (hatched) for the three types of fast
  emission compared with observations (see text for details). }
\label{exotica}
\end{figure}

After several examples we can turn to a general analysis.

One can show that it is sufficient to lower
the neutrino luminosity due to the modified Urca
process in a low-mass star by a factor of $\sim 30-100$
to explain observations of stars hottest for their age.
This low neutrino luminosity can be produced by
nucleon-nucleon bremsstrahlung. The modified Urca
process can be reduced by nucleon superfluidity,
as in the above examples (Figs.\ \ref{cool4}c and d).

The level of fast neutrino emission in high-mass stars
is less certain. To analyze it, let us
consider (Fig.\ \ref{exotica}) three typical levels
appropriate to the direct Urca process in nucleon/hyperon
matter; to pion-condensed matter; and kaon-condensed
matter. A schematic picture of the neutrino emissivity
profiles through the stellar core for these scenarios
is presented in Fig.\ \ref{exotica}a. 

Fig.\ \ref{exotica}b
shows the ranges of $T_{\rm s}^\infty$ 
(hatched regions)
which can be explained
by the theory assuming the scenario with the direct
Urca process. The upper curve shows cooling of low-mass
stars via neutron-neutron bremsstrahlung neutrino emission
(as in Figs.\ \ref{cool4}c and d). The lower curve presents
fast cooling of a maximum-mass star with direct Urca
process open in the inner core. Any value  of $T_{\rm s}^\infty$
between these two curves can be explained by a cooling
of a star with some intermediate mass. As discussed above,
this scenario explains all the data
and predicts the existence of very cold neutron stars.

Fig.\ \ref{exotica}c refers to the scenario with
pion-condensation. The upper cooling
curve is the same as in Fig.\ \ref{exotica}b
(because low-mass stars have no 
inner cores), but the lower curve goes essentially
higher (because the neutrino emission due to pion
condensation is lower than due to direct Urca process).
This scenario also explains all the sources and
predicts the existence of cold stars, but not so cold
as in the direct Urca case.

Fig.\ \ref{exotica}d is a similar plot for the scenario
with kaon-condensation in the inner core. 
High-mass stars cool slower than in the previous scenarios,
but all the data are still explained and a population
of stars slightly colder than the observed ones is predicted.

Thus, all three scenarios are currently consistent with
the observations, but predict different families 
of cold stars. Similar conclusions have been made
by a number of authors (see, e.g., \cite{yp04,pgw06}
and references therein).
Calculations show that the
the neutrino luminosity in the inner core should be
$\approx$30--100
 times higher than the modified Urca luminosity 
to explain all the sources and leave
no space for neutron stars colder than the observed ones.
A discovery of very cold neutron stars would be
crucial to firmly constrain the enhanced
neutrino emission level.

Unfortunately, current observations give no reliable
constraints on the position of the transition layer
between slowly and fastly neutrino emitting regions in the
core (on $\rho_1$ and $\rho_2$ or $M_1$ and
$M_2$ in Fig.\ \ref{concept}).

\section{Other aspects}

We do not discuss in detail many aspects of the
cooling theory  
but mention them here.

First, cooling can also be regulated
by heat conduction in the neutron star crust. 
The thermal conductivity in the crust can be
affected by composition of the matter
(e.g., by the presence of light elements in accreted surface layers)
and magnetic fields (in the outer and inner
crust; see, e.g., \cite{yp04,pgw06,gkp06}
and references therein). However, these cooling regulators
are usually not as strong as neutrino emission from the core.

The effects of superfluidity can be more sophisticated
than described above. 
Taking different 
EOSs [different direct Urca thresholds
and $T_{\rm c}(\rho)$ profiles]
one can obtain a complicated zoo of families
of cooling neutron stars \cite{kaminkeretal06a, gkyg05}.
Neutrino emission due to Cooper
pairing of baryons can affect the cooling in many ways
as reviewed in Refs.\ \cite{yp04,pgw06}. In particular,
a moderately strong neutron superfluidity in the nucleon core
can enhance the neutrino luminosity over the modified
Urca level by a factor of $\sim$30, allowing one to
explain the coldest observed sources by neutrino emission
from the nucleon (non-exotic) core without invoking the direct
Urca process. This leads to the scenario of ``minimal cooling''
\cite{pageetal04,gusakovetal04}. 

Cooling of neutron stars can be accompanied by some
reheating associated, for instance, with the dissipation
of differential rotation, deviation from
beta-equilibrium or decay of magnetic fields
in neutron star interiors. The 
magnetic reheating can be especially strong in magnetars
(see below). Otherwise these reheating mechanisms
are expected to be significant in old neutron stars
(at the photon cooling stage). They are reviewed in
\cite{yp04,pgw06}.

Finally, we have not discussed cooling of strange
stars and neutron stars with quark cores. A comprehensive 
review is given in \cite{pgw06}.

\section{Connections}

 With the spectacular
progress of neutron star observations, cooling problem becomes
more connected to other aspects of neutron star physics.

First, cooling of isolated neutron stars is
closely related to deep crustal heating of transiently
accreting neutron stars in soft X-ray transients (SXTs).
The matter accreted on a neutron star becomes eventually
compressed by the weight of newly accreted material and
sinks thus into the crust. Sinking into
the deep crust is accompanied by nuclear transformations
\cite{hz90,hz03,guptaetal07,hz07} 
and associated heating of the entire star ($1.5-2$ MeV per
one accreted nucleon).
This deep crustal heating can keep 
thermally inertial stars warm \cite{bbr98} even in quiescent states
of SXTs when accretion is switched off.
It is likely responsible for
thermal emission observed from some SXTs
in quiescent states (e.g., Refs.\ \cite{heinke07,jonker07}). 

One can distinguish two heating regimes.
As a rule, accretion episodes are
not too long (weeks-months) and do not violate
isothermality of internal layers. These stars
reach thermal quasiequilibrium determined by
the mass accretion rates $\langle \dot{M} \rangle$
averaged over their global thermal relaxation time 
($\sim 100-1000$ yr). 
Observations of these objects test essentially the same
physics of neutron star interior 
as observations of cooling isolated 
neutron stars (e.g., \cite{ylh03,yak-ea04,lh07}).
The analysis of the data on SXTs \cite{lh07,heinke07,jonker07}
generally agrees with the above analysis of cooling
neutron stars, but with important exceptions.
The upper limits on $T_{\rm s}$ in quiescent states
of the transiently accreting millisecond pulsar SAX J1808.4--3658
and possibly the SXT 1H 1905+000
are very low, giving an example of very cold neutron
stars (missing so far in the data on cooling neutron stars).
If true, these neutron stars undergo strong neutrino
cooling via direct Urca process. 
However, in view
of uncertainties associated with the deep crustal heating
hypothesis, determination of 
$T_{\rm s}$ and $\langle \dot{M} \rangle$, these results
should be taken with caution.

Some SXTs (KS~1731--260 and MXB 1659--29, e.g.,
\cite{cackettetal06}) 
show long accretion episodes (a few years and more)
in which the crust may become much warmer than the
core. After accretion stops, the crust cools and thermally
equilibrates with the core. This crust-core relaxation is
reflected in the relaxation of the surface temperature $T_{\rm s}(t)$,
that is observed and can give important information on
the  crust and core physics
\cite{ur01,rutledgeetal02,cackettetal06,shterninetal07}.  

Cooling theory is also employed to study thermal states
of magnetars. It seems one needs to introduce a reheating
(most probably produced by strong magnetar magnetic fields)
to explain high observed X-ray luminosities of magnetars.
One can study quasistationary thermal states of magnetars
(e..g., \cite{kaminkeretal06b}) and afterburst relaxation
(e.g., \cite{lubarskyetal}).     

In addition, cooling theory can be used to analyze
the propagation of thermal waves in neutron stars
during X-ray bursts and superbursts
(e.g., \cite{fujimotoetal84,cummingetal06})
and to explore cooling of vibrating neutron stars
\cite{gyg05}. 

Since the data are becoming numerous it is possible
to combine cooling theory with neutron star 
statistics \cite{popovetal06}.

\section{Conclusions}

The theory of cooling neutron stars of ages
$10^2-10^6$ yr mostly tests the neutrino emission
properties of the neutron star core. Its main results
are as follows. 

(1) Neutrino emission in the
outer core (i.e., in the core of a low-mass star)
is a factor of 30--100 lower than the
modified Urca emission in a nonsuperfluid star. 

(2) Neutrino emission in the inner core (of a massive
star) is at least a factor of 30--100 higher than
the modified Urca emission. It can be enhanced by
direct Urca process in nucleon/hyperon inner core
or by the presence of pion or kaon condensate,
or quark matter.

(3) The scenario with open direct Urca process 
predicts the existence (Fig.\ \ref{exotica})
of massive isolated neutron stars which are much
colder than those observed now. In the scenario with
pion condensate, the massive stars should be warmer
(than those with open direct Urca) but colder than
the observed ones. In the scenario with kaon condensate
the massive stars should be even warmer but
slightly colder than the observed sources.
A discovery of cold cooling neutron stars would
be crucial to constrain the level of enhanced neutrino
emission in the inner core.

(4) Observations of cooling neutron stars can be
analyzed together with observations of SXTs in quiescent
states. The data on SXTs indicate the existence of
very cold neutron stars (first of all, SAX J1808.4--3658) 
which cool via direct Urca process, but the data and
interpretation require additional confirmation.

(5) A transition from slow
neutrino emission in the outer core to enhanced emission
in the inner core has to be smooth. Current observations
of cooling neutron stars and SXTs do not 
constrain the parameters of this transition.
A firm measurement of masses of cooling
or accreting stars would 
help to impose such constraints. 
 
(6) New observations and reliable practical theories
of dense matter are vitally important to tune the cooling theory
as an instrument for exploring physical properties of
neutron star interiors and neutron star parameters.  
The tuning will imply a careful analysis of many 
cooling regulators.


\begin{theacknowledgments}
This work was partially supported  by 
the Russian Foundation for Basic Research
(grants 05-02-16245 and 05-02-22003),
by FASI-Rosnauka (grant NSh 9879.2006.2),
and by the Joint Institute for Nuclear Astrophysics
(grant NSF PHY 0216783).
\end{theacknowledgments}



\bibliographystyle{aipproc}   




\end{document}